# A Test to Verify the Change of Light's Speed in the Gravitational Field of the Earth

## ----A New Verification of Experiment for General Relativity

Mei   Xiaochun

(Department of Physics, Fuzhou University, China, E-mail: mxc001@163.com)

**Abstract** Based on the Schwarzschild solution of the Einstein's equation of gravitational field, it is proved that the speed of light would change and the isotropy of light's speed would be violated in gravitational field with spherical symmetry. On the earth's surface, the speed of light vertical to the surface is $0.2 \text{m/s}$ less than that parallel to the surface. It is suggested to use the method of the Michelson—Morley interference to verify the change of light's speed and the violation of isotropy in the gravitational field of the earth. In the proposed experiment, one arm of interferometer is vertical to the earth's surface while another is parallel to the earth's surface so that the difference of gravitational potential exists between them. When two arms are turned over $90^0$, the shift of about $0.07$ interference stripe would be caused which can be observed directly. So this experiment can be considered as a new verification for general relativity in the gravitational field with spherical symmetry. (In the past experiments of the Michelson—Morley interference, the interferometer's two arms were arranged to be parallel to the earth's surface without the difference of gravitational potential between them. In this case, the shift of interference stripe could not be observed.) If the experiment shows that gravitation would change light' speed and violate the isotropy of light's speed, the result would cause great effects on foundational physics, astrophysics and cosmology.

In light of special relativity, light's speed is a constant with isotropy in vacuum without the existence of gravitation or interaction. But it is still an uncertain problem in the current theory whether or not light's speed would change in gravitational field. Because there is interaction between photon and material in gravitational field, photon would not be free one. It is a rational speculation that light's speed would change in gravitational field. As shown below, we prove this point based on the Einstein's theory of gravitation. On the other hand, according to the common understanding of general relativity, we have both definitions of the coordinate time and the proper time, as well as the coordinate length and the proper length. Proper time and proper length are also called as stander clock and stander ruler. Because coordinate time and coordinate length are variable in the different space-time points in gravitational field, it is meaningless for us to use them do any measurement. When we measure time, space and object's motions in gravitational field, what can be used are only stander clock and stander ruler according to the current theory. It has been proved that in the static gravitational field, when stander clock and stander ruler are used, light's speed, to be equal to its speed in vacuum, is still a constant[1]. Therefore at present, in the concrete calculations of astrophysical and cosmological problems, as well as in the theoretical explanations of astronomical observations, we always take light's speed as a constant.

However, these two judgments are contradictory each other. Which one is real? To resolve this problem, we concretely discuss the speed of light in the static gravitational field with spherical symmetry. Meanwhile, an experiment is proposed for the observers who are at rest in gravitational field to judge whether or not light's speed would change in the field. The Schwarzschild metric of the Einstein's equation of static gravitational field with spherical symmetry is

$$ds^2 = c^2\left(1-\frac{\alpha}{r}\right)dt^2 - \left(1-\frac{\alpha}{r}\right)^{-1}dr^2 - r^2\left(\sin^2\theta d\varphi^2 + d\theta^2\right) \quad (1)$$

In which $\alpha = 2GM/c^2$, $dt$ is coordinate time, $dr$, $d\theta$ and $d\varphi$ are coordinate lengths (angles). For light's motion, we have $ds = 0$. Let $\tau$ represent proper time and take $\theta = \pi/2$. Then substitute (1) into the equation of geodetic line, we can get integrals

$$c\left(1-\frac{\alpha}{r}\right)\frac{dt}{d\tau} = \varepsilon \qquad r^2\frac{d\varphi}{d\tau} = L \quad (2)$$

Here $\varepsilon$ and $L$ are constants. $L$ is actually the unit angle momentum of photon in gravitational field. Let $\varepsilon = 1$, we get

$$d\tau = \left(1-\frac{\alpha}{r}\right)dt \quad (3)$$

So (1) can be written as

$$\left(1-\frac{\alpha}{r}\right)\left(\frac{dt}{d\tau}\right)^2 - \left(1-\frac{\alpha}{r}\right)^{-1}\left(\frac{dr}{cd\tau}\right)^2 - r^2\left(\frac{d\varphi}{cd\tau}\right)^2 = 0 \quad (4)$$

or

$$\left(\frac{dr}{d\tau}\right)^2 = c^2\left[1-\left(1-\frac{\alpha}{r}\right)\frac{L^2}{c^2 r^2}\right] \quad (5)$$

By taking differential about $d\tau$ in (5), we get

$$\frac{d^2r}{d\tau^2} - \frac{L^2}{r^3} = -\frac{3\alpha L^2}{2r^4} \quad (6)$$

By comparing with the Newtonian formula of gravitation in the reference frame of polar coordinate

$$\frac{d^2r}{dt^2} - r\left(\frac{d\varphi}{dt}\right)^2 = -\frac{c^2\alpha}{2r^2} \qquad r^2\frac{d\varphi}{dt} = L \quad (7)$$

we can write (6) as

$$\frac{d^2\vec{r}}{d\tau^2} = -\frac{3\alpha L^2 \vec{r}}{2r^5} \quad (8)$$

By using (3) and relation $\vec{V} = \vec{V}_r + \vec{V}_\varphi = V_r\vec{e}_r + V_\varphi\vec{e}_\varphi$, the formula (8) can also be written as

$$\frac{d^2\vec{r}}{dt^2} = -\left(1-\frac{\alpha}{r}\right)^2 \frac{3\alpha L^2 \vec{r}}{2r^5} + \left(1-\frac{\alpha}{r}\right)\frac{\alpha V_r \vec{V}}{r^2} \quad (9)$$

From (3) and (5), we obtain light's speeds in the gravitational field with spherical symmetry

$$V_r^2 = \left(\frac{dr}{dt}\right)^2 = \left(\frac{dr}{d\tau}\frac{d\tau}{dt}\right)^2 = c^2\left(1-\frac{\alpha}{r}\right)^2\left[1-\left(1-\frac{\alpha}{r}\right)\frac{L^2}{c^2r^2}\right] \qquad (10)$$

$$V_\varphi^2 = \left(r\frac{d\varphi}{dt}\right)^2 = \left(r\frac{d\varphi}{d\tau}\frac{d\tau}{dt}\right)^2 = \left(1-\frac{\alpha}{r}\right)^2\frac{L^2}{r^2} \qquad (11)$$

$$V = \pm\sqrt{V_r^2 + V_\varphi^2} = \pm c\left(1-\frac{\alpha}{r}\right)\sqrt{1+\frac{\alpha L^2}{c^2r^3}} \qquad (12)$$

It is obvious that $V$ is not only relative to the mass of center material, but also relative to the angular momentum of photon. It can be seen from (8) and (9) that light's accelerations are not equal to zeros, no matter what proper time or coordinate time is used. So light's speed can not be a constant. Meanwhile, because it is relative to angular momentum, light's speed is also anisotropic in gravitational field.

It should be noted that $d\tau$ is the proper time indicated by the clock fixed on the moving reference frame, and $dt$ is the coordinate time indicated by the clock located on the static reference frame. When formula (1) is used to describe the earth's gravitational field, $dt$ and $dr$ in (9) are actually the time and length which are measured by the observers at rest on the earth. So in light of (12), for observers who are at rest on the earth's surface, light's speed is not a constant and not be isotropic again. On the other hand, as we known that only on the locally inertial reference frame, we can define stander ruler (proper length) and stander clock (proper time). The locally inertial reference frame is considered to be one falling freely in gravitational field, in which the Schwarzschild metric becomes that of flat space-time with the form $ds^2 = c^2d\tau^2 - dR^2$. So only on the locally inertial reference frame with stander ruler and stander clock, light's speed would be a constant equal to its speed in vacuum. But for observers at rest in the gravitational field of the earth which is not a locally inertial reference frame, light's speed could not be a constant, because no stander ruler and stander clock can be defined according to the current understanding.

In fact, in general relativity, when we calculate the perihelion precession of the Mercury, the deviation of light and the delay of radar wave in the solar gravitational field, we actually use coordinate time and coordinate length, in stead of proper time and proper length. The calculating results of the deviation of light and the delay of radar wave in the gravitational field of the sun are

$$\theta = \frac{4GM}{c^2 r_s} \qquad \Delta t = \frac{4GM}{c^3}\left(1+ln\frac{4rr'}{r_s}\right) \qquad (13)$$

Here $r_s$ is the solar radius, $r$ and $r'$ are the distances between the earth, the Mercury and the sun. It can be seen from the deduction processes of (13) that the space-time coordinates $r, r', r_s, \theta$ and $t$ in the formula are the same as defined in formula (1). By comparing the formula (13) with experiments directly, we claim that general relativity is supported by experiments. However, we should be clear in mind that the description of (13) is based on the static reference frame of the sun, in which coordinate time and coordinate length are used. We have not transformed them into proper time and proper length. Meanwhile, it should be noticed that all experiments and observations are carried out on the earth reference frame, in stead of that on the sun. Only because relative velocity between them is small, the effect of special relativity is neglected.

Therefore, for the same problem about light's motion in gravitational field, it may be possible for us to verify that the earth's gravitational field would affect the light's speed and violate its isotropy through experiments. The method of the Michelson—Morley interference is suggested for this purpose. When light moves along the vertical direction of the earth's surface, we have $L = 0$. So according to (12), light's vertical speed is

$$c_\perp = c\left(1 - \frac{\alpha}{r}\right) \qquad (14)$$

Here $c$ is light's speed when $r \to \infty$. When light moves along the parallel direction of the earth's surface, we have $L = cr_e$. Here $r_e$ is the earth's radius. We have $\alpha/r_e = 1.39 \times 10^{-9} \ll 1$ for the earth. So according to (12), light's parallel speed on the earth's surface is

$$c_{11} = c\left(1 - \frac{\alpha}{r_e}\right)\sqrt{1 + \frac{\alpha}{r_e}} \approx c\left(1 - \frac{\alpha}{2r_e}\right) \qquad (15)$$

If let $V_r = 0$ in (10), we get $L/(cr) = (1 - \alpha/r)^{-1/2}$. Substitute the result into (11), we can also get the same result. At present, light's speed is generally measured on the earth's surface, according to the current measurement, we can take $c_{11} = 2.997924580 \times 10^8 m/s$. Therefore, on the earth's surface, we have

$$c_\perp = \frac{c_{11}}{\sqrt{1 + \alpha/r_e}} \approx c_{11}(1 - 6.95 \times 10^{-10}) = 2.997924578 \times 10^8 m/s \qquad (16)$$

So we have $c_{11} - c_\perp = 0.2 m/s$. It means that the light's vertical speed is less than its parallel speed on the earth's surface. Thought the change of light's speed and the violation of isotropy are very small, we can still verify their existence by using the method of the Michelson—Morley interference. Suppose that the Michelson interferometer's two arms have the same length $h$. One arm is vertical to the earth's surface and another is parallel to the surface. In this way, there exists the difference of gravitational potential between them. When a bundle of light moves along the vertical direction of the earth's surface from the point $r_e$ to the point $r_e + h$ with $h \ll r_e$, spent time is

$$\Delta t_1 = \int_{r_e}^{r_e+h} \frac{dr}{c_\perp} = \int_{r_e}^{r_e+h} \frac{dr}{c(1-\alpha/r)} = \frac{1}{c}\left[h + \alpha \ln\left(1 + \frac{h}{r_e - \alpha}\right)\right] \approx \frac{h}{c}\left(1 + \frac{\alpha}{r_e}\right) \qquad (17)$$

While a bundle of light moves $h$ distance along the parallel direction of the earth's surface, spent time is

$$\Delta t_2 = \frac{h}{c_{11}} \approx \frac{h}{c}\left(1 + \frac{\alpha}{2r_e}\right) \qquad (18)$$

So when two lights are reflected back, the time difference is

$$\Delta T = 2(\Delta t_1 - \Delta t_2) = \frac{\alpha h}{c r_e} \tag{19}$$

When the interferometer's arms are turned over $90^0$, the time deference becomes

$$\Delta T' = -2(\Delta t_1 - \Delta t_2) = -\frac{\alpha h}{c r_e} \tag{20}$$

So before and after the interferometer's arms are tuned, the change of time difference is

$$\delta t = \Delta T - \Delta T' = \frac{2\alpha h}{\lambda r_e} \tag{21}$$

We take light's wave length $\lambda = 4 \times 10^{-7} m$ and $h = 10 m$ in the experiment. After interferometer's arms are tuned over $90^0$, the shift of interference stripe would be

$$\Delta = \frac{c}{\lambda} \delta t = \frac{2\alpha h}{\lambda r_e} = 6.95 \times 10^{-2} \tag{22}$$

Because the shift of $10^{-2}$ stripe can be observed for the precise Michelson interferometer[2], this is a directly observable result. If the shift of interference stripe can be found, it indicates that light's speed would change and its isotropy would be violated in the gravitational field of the earth. In the past experiments of the Michelson—Morley interference, which were used to prove light's speed invariable were, the interferometer's two arms were arranged to be parallel to the earth's surface without the existence of gravitational potential's difference. This may be one of the reasons that the shift of interference stripe could not be found. If there is the difference of gravitational potential between two arms, the result would be different. It would be possible for us to observer the shift of interference stripe.

**Discussion** Because the formula (12) is deduced from the Schwarzschild solution of the Einstein's equation of gravitational field, the experiment can be regarded as a new verification for general relativity in the weak gravitational field. It should be emphasized that all three theoretical predictions shown in the formulas (13) and (22) are deduced from the formula (4). Now that the effects shown in (13) have been verified to exist, the effect shown in (22) would also exist for the observers at rest in the gravitational field. Thought the change of light's speed in the earth's gravitational field is very small, it would be great in strong field. For example, we have $\alpha / r = 2.12 \times 10^{-6}$ on the surface of the sun. So for the observation on the sun's surface, we would have $c_{11} - c_\perp \approx 3 \times 10^2 \, m/s$. This is quite big value though the gravitational field of the sun is still not very strong. Under the limit condition, on the so-called black hole's surfaces with $\alpha / r \to 1$, we would have light's speed $V \to 0$ in light of (12). But according to the current understanding, in this case we still have $\lambda \nu = c \neq 0$ with light's wave length $\lambda \to \infty$ and frequency $\nu \to 0$. These two results are completely different. At present, light's speed is always regarded as a constant in the concrete calculations of astrophysical and cosmological problems, as well as in the theoretical explanations of astronomical observations. If the experiment shows that light's speed would change in gravitational field, the result would produce great effects on foundational physics, astrophysics and cosmology. Conversely, if the shift of interference stripe can not be observed in the experiment, we should ask why the prediction of general relativity can not coincide with practical observations. So no matter whether or not the shift of interference stripe can be observed, the experiment is well meaningful and worthy to be done seriously.